# Low temperature transport on surface conducting diamond

M. T. Edmonds[1*], L. H. Willems van Beveren[2], K. Ganesan[2], N. Eikenberg[2], J. Cervenka[2], S. Prawer[2], L. Ley[1], A. R. Hamilton[3], C. I. Pakes[1]

[1]Physics Department, La Trobe University, Victoria 3086, Australia
[2]School of Physics, University of Melbourne, Victoria 3010, Australia
[3]School of Physics, University of New South Wales, New South Wales 2052, Australia
*Corresponding author: Email mtedmonds@students.latrobe.edu.au

**Abstract:** Magneto-transport measurements were performed on surface conducting hydrogen-terminated diamond (100) hall bars at temperatures between 0.1-5 K in magnetic fields up to 8T.

**1 Introduction:** Diamond with a band gap of 5.47 eV is an insulating material when undoped. However, when the diamond surface is terminated with hydrogen and then exposed to air or synthetic surface adsorbates, a *p*-type surface conductivity is observed [1].

Changes in the electronic properties of the diamond surface as a result of this p-type surface conductivity are well understood [2,3], and transport measurements on surface conducting diamond have been performed at room temperature and liquid nitrogen temperatures [4,5]. However, little attention has been paid to the transport properties of the p-type carriers in diamond at liquid helium temperatures and in high magnetic fields.

**2 Experiment:** Two samples were used in this study: an electronic grade sample (E1) with low impurity content and surface roughness ~0.1 nm and one standard grade sample (SG1) with a higher level of impurities and larger surface roughness ~1.0 nm. Hall Bars were fabricated on these samples using standard photolithography techniques. In the inset of Fig. 1 the blue region of the Hall Bar represents the h-terminated conducting region and the region outside of the Hall Bar is insulating oxygen-terminated diamond. After fabrication magneto-transport measurements were performed using a dilution refrigerator at temperatures between 0.1-5.0 K in magnetic fields up to 8 T.

**3 Results:** Figure 1 shows the sheet conductivity as a function of temperature between room temperature and 0.1K. Both samples show only weak temperature dependence with the electronic grade sample having a higher room temperature sheet conductivity. Carrier activation energies have been determined from Fig. 1 to be 0.43 meV and 0.14 meV for SG1 and E1 respectively in the linear region.

Longitudinal resistivity $\rho_{xx}$, and Hall resistivity $\rho_{xy}$ was measured as a function of magnetic field perpendicular to the sample, for temperatures between 0.3 K and 5.0 K for both samples. Fig. 2 shows the data for E1 at 3.9 K. From a linear fit to the Hall resistivity as a function of magnetic field the hole concentration is found to be ~$10^{13}$ cm$^{-2}$, very close to that obtained at room temperature indicating no carrier freeze out. The mobility of 78 cm$^2$V$^{-1}$s$^{-1}$ has slightly decreased from the room temperature value of 105 cm$^2$V$^{-1}$s$^{-1}$; this follows the change in sheet conductivity which decreases with temperature. This low carrier mobility at cryogenic temperature is attributed to scattering from adsorbate anions.

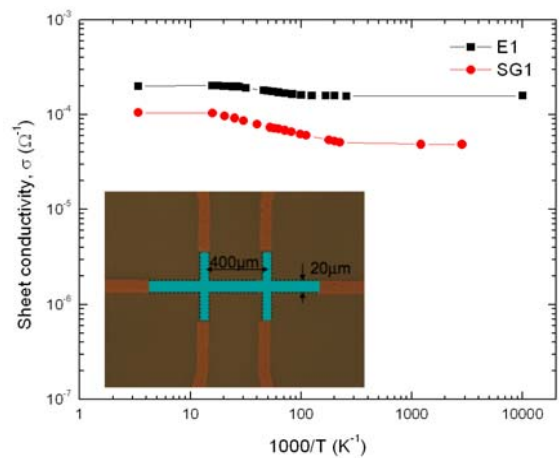

Fig. 1. Main: Sheet conductivity as a function of temperature. Inset: Image of a Hall Bar fabricated on IIa (100) diamond.

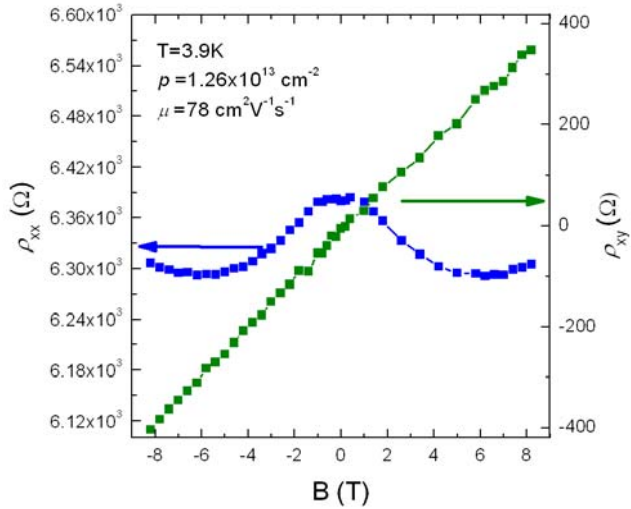

Fig. 2. Longitudinal resistivity $\rho_{xx}$, and Hall resistivity $\rho_{xy}$ versus B at 3.9 K for E1.

Surface conductivity on diamond deals with a 2D system with carriers and scatterers confined to two adjacent planes separated by only a few nanometers [2]. With such a small separation between the charge sheets, scattering from anions in the adsorbed water layer is expected to be the dominant scattering mechanism.

Magneto-resistance as a function of magnetic field in Fig. 2 exhibits weak localization similar to that observed for highly disordered Si:P δ-doped two-dimensional electron systems [6].

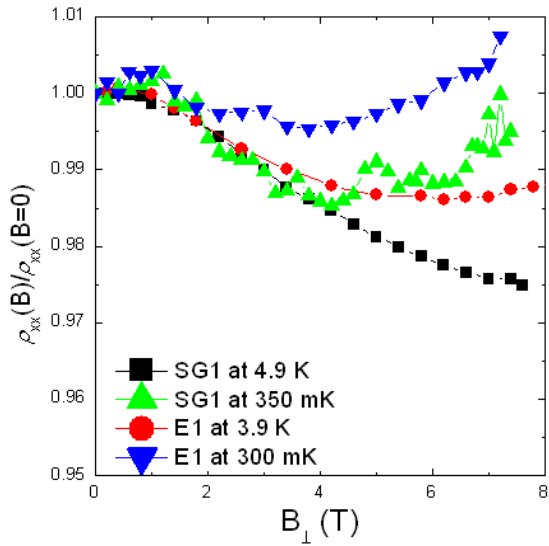

Fig. 3. Normalized magneto-resistance for SG1 and E1 versus B for temperatures between 0.3 K and 5.1 K.

Fig. 3 shows normalized magneto-resistance data at two different temperatures for both samples. The data taken at sub-0.4 K display oscillations that require further investigation but are more likely related to changes in sample temperature than any quantum behaviour.

Further magneto-transport measurements are required between 4K and mK at constant temperature to fully understand the hole-hole interaction in this highly disordered system and to obtain the weak localization correction to the Drude conductivity

## Acknowledgements

This work has been supported by the Australian Research Council under DP0879827.

## References

[1] F. Maier, M. Riedel, B. Mantel, J. Ristein, and L. Ley, "Origin of surface conductivity in diamond," Phys. Rev. Lett. 85, 3472 (2000)

[2] M. T. Edmonds, C. I. Pakes, S. Mammadov, W. Zhang, A. Tadich, J. Ristein, and L. Ley, "Surface band bending and electron affinity as a function of hole accumulation density in surface conducting diamond," Appl. Phys. Lett. 98, 102101 (2011)

[3] M. T. Edmonds, M. Wanke, A. Tadich, H. M. Vulling, K. J. Rietwyk, P. L. Sharp, C. B. Stark, Y. Smets, A. Schenk, Q.-H. Wu, L. Ley, and C. I. Pakes, "Surface transfer doping of hydrogen-terminated diamond by $C_{60}F_{48}$: Energy level scheme and doping efficiency" J. Chem. Phys. 136, 124701 (2012)

[4] J. A. Garrido, T. Heimbeck, and M. Stutzmann, "Temperature-dependent transport properties of hydrogen-induced diamond surface conductive channels" Phys. Rev. B 71, 245310 (2005)

[5] C. E. Nebel, C. Sauerer, F. Ertl, M. Stutzmann, C. F. O. Graeff, P. Bergonzo, O. A. Williams, and R. Jackman, "Hydrogen-induced transport properties of holes in diamond surface layer" Appl. Phys. Lett. 79, 4541 (2001)

[6] K. E. J. Goh, M. Y. Simmons, and A. R. Hamilton, "Electron-electron interactions in high disordered two-dimensional systems" Phys. Rev. B 77, 235410 (2008)